\documentclass{emulateapj}

\usepackage{amsmath,natbib}

\usepackage{url}
\usepackage[backref,breaklinks,colorlinks,citecolor=blue]{hyperref}
\usepackage[all]{hypcap}

\usepackage{color}
\definecolor{darkblue}{rgb}{0.0, 0.0, 0.5}
\hypersetup{colorlinks=true,citecolor=blue,linkcolor=darkblue}

\newcommand{\tb}{t_\bot}
\newcommand{\tp}{t_\parallel}
\newcommand{\id}{i_{disk}}
\newcommand{\pd}{\phi_{disk}}

\newcommand{\taur}{\tau(r)}

\newcommand{\actaa}{Acta Astronomica}
\newcommand{\icarus}{Icarus}

\slugcomment{ApJ accepted 2014 Dec 28}

\shorttitle{Exorings}
\shortauthors{Kenworthy and Mamajek}

\begin{document}

\title{Modeling giant extrasolar ring systems in eclipse \\
and the case of J1407b: sculpting by exomoons?}

\author{M.A. Kenworthy}
\affil{Leiden Observatory, Leiden University, P.O. Box 9513, 2300 RA Leiden}

\and

\author{E.E. Mamajek}
\affil{Department of Physics and Astronomy, University of Rochester,
Rochester, NY 14627-0171, USA}

\begin{abstract}

The light curve of 1SWASP J140747.93-394542.6, a $\sim$16 Myr old star
in the Sco-Cen OB association, underwent a complex series of deep
eclipses that lasted 56 days, centered on April 2007 \citep{Mamajek12}.
This light curve is interpreted as the transit of a giant ring system
that is filling up a fraction of the Hill sphere of an unseen secondary
companion, J1407b \citep{Mamajek12,vanWerkhoven14,Kenworthy15}.
We fit the light curve with a model of an azimuthally symmetric ring
system, including spatial scales down to the temporal limit set by the
star's diameter and relative velocity.
The best ring model has 37 rings and extends out to a radius of 0.6 AU
(90 million km), and the rings have an estimated total mass on the order
of $100 M_{Moon}$.
The ring system has one clearly defined gap at 0.4 AU (61 million km),
which we hypothesize is being cleared out by a $< 0.8 M_{\oplus}$
exosatellite orbiting around J1407b.
This eclipse and model implies that we are seeing a circumplanetary disk
undergoing a dynamic transition to an exosatellite-sculpted ring
structure and is one of the first seen outside our Solar system.

\end{abstract}


\section{Introduction}

Circumstellar disks of gas and dust are a ubiquitous feature of star
formation.
Circumstellar gas-rich disks disperse on timescales of $<$10 Myr,
effectively limiting the runaway growth phase for gas giant planets
\citep{Williams11}.
The architecture of the resultant planetary systems is dictated by the
structure and composition of the disk, its interaction with the young
star, and the competing formation mechanisms that transfer
circumstellar material onto accreting protoplanets
\citep[e.g. see reviews by][]{Armitage11,Kley12}. 
Extended month- to year-long eclipses indicate the presence of long
lived dark disks around secondary companions, including $\epsilon$
Aurigae \citep{Guinan02, Kloppenborg10}, EE Cep \citep{Mikolajewski99,
  Graczyk03, Mikolajewski05}, a precessing circumbinary disk around KH
15D \citep{Hamilton05,Winn06b} and three systems recently discovered
in the OGLE database -- OGLE-LMC-ECL-17782 \citep{Graczyk11},
OGLE-LMC-ECL-11893 \citep{Dong14} and OGLE-BLG182.1.162852
\citep{Rattenbury14}.

Gas planets are thought to form through accretion from circumstellar disks
composed of gas and dust.
Angular momentum of the circumstellar disk material is redistributed
through the formation of a circumplanetary disk.
After the gas is cleared out of the planetary system, dust in the
circumplanetary disk then accretes into moons or remains as a ring
system within the Roche limit of the planet \citep{Canup02, Magni04,
  Ward10}.
The transits of giant planets with ring systems produces a distinct
and detectable light curve \citep{Barnes04, Tusnski11}, and searches
for the transit timing variations caused by attendant exomoons are
ongoing \citep{Kipping12, Kipping13}.
1SWASP J140747.93-394542.6 (hereafter J1407) is a pre-main sequence
$\sim 16$ Myr old star, $0.9 M_{\odot}$, $V=12.3$ mag K5 star at 133
pc associated with the Sco-Cen OB Association \citep{Mamajek12,
  vanWerkhoven14,
  Kenworthy15}\footnote{\url{http://exoplanet.eu/catalog/1swasp_j1407_b/}}.
The Super Wide Angle Search for Planets (SuperWASP) database
\citep{Butters10} showed that the star underwent a complex series of
eclipses lasting $\sim56$ days around May 2007 and including a dimming
of $ > 95 \%$.
\citet{Mamajek12} and \citet{vanWerkhoven14} propose that these
eclipses are caused by a large ring system orbiting an unseen
substellar companion, dubbed J1407b.
In the first attempt to model the system using nightly averaged
photometry, \citet{Mamajek12} posited at least four large rings
girding J1407b.
A more detailed analysis of the SuperWASP raw data by
\citet{vanWerkhoven14} and removal of a 0.1 mag amplitude, 3.2 day
periodic variability due to rotational modulation by star spots shows
temporal structure down to a limit of 10 minutes
\citep{vanWerkhoven14}.
Only the 2007 eclipse event is seen in the time series photometry, and
\citet{Kenworthy15} place constraints on the possible mass and orbital
period for this companion.
They conclude that J1407b is almost certainly substellar (at
$>$3$\sigma$ significance), and possibly an exoplanet.

The analysis of eclipse light curves to determine the structure of
otherwise unresolved astrophysical objects is possible for specific
cases.
Since the first proposal in \citet{MacMahon08}, high speed photometry of
lunar occultations has been used \citep{White87} to determine the
multiplicity of stars close to the Ecliptic, structure of evolved stars
and studies of the Galactic centre.
The sharp inner ring edge of Saturn's Encke gap has been used to deduce
the wavelength-dependent radius of Mira using high speed photometry from
{\it Cassini} \citep{Stewart13}.
The inverse problem of determining the extended structure of a
foreground object assuming a point-like background source has been
used to discern fine structures in the rings of the gas giant planets,
measure the scale heights of the atmospheres in planetary bodies such
as Titan and Pluto \citep[e.g.][]{McCarthy08}, and to determine the
shape and orbital properties of Solar system asteroids
\citep[e.g.][]{Dunham90,Shevchenko06}.
Most recently, two rings were discovered around an asteroid in the Solar
system \citep{BragaRibas14}, where the structure in the rings themselves
are unresolved.
In this paper, we use knowledge of an extended background source and
multiple ring structure around a foreground object to derive the
geometric properties of the J1407b ring system, a candidate
circumplanetary disk.

In Section \ref{sec:model} we present our exoring model and then discuss
several features to be expected in a ring system transit where the ring
system is significantly larger than the parent star.
In Section \ref{sec:j1407} we present our best fits to the J1407b
transit data, and in Section \ref{sec:structure} we discuss the structure in the most
plausible ring models and we posit that they are indirect evidence for
exomoons or exosatellites coplanar with the rings.
The large size of the ring system presents a challenge to what type of
orbit it must have around the primary star, which we address in Section
\ref{sec:orbit}.
Our conclusions are presented in Section \ref{sec:conc}.

\section{Ring Model}
\label{sec:model}

Our ring model is composed of two parts, which we solve sequentially for
an observed transiting ring system.
We assume that the primary star and ring system are at a similar
distance from the Earth, and that the ring system is at least several
times larger than the angular size of the primary star.
We first solve for the orientation of the plane of the ring system
relative to our line of sight, and we then solve for the transmission of
the rings as a function of radius from the secondary companion, given
the geometry of the ring system derived in the previous step.

\subsection{Input parameters}

The primary is a star at a distance of $d$ parsecs, with radius
$R_\star$.
We approximate the orbit of the secondary for the duration of the
eclipse as being a straight line, with constant relative velocity of
$v$.
Surrounding the secondary companion is a ring system in a plane that
contains both the rings and the equatorial plane of the secondary
companion, which we refer to as the ring plane.
The rings are composed of individual particles that orbit the
secondary companion in Keplerian orbits, assumed circular.
These particles scatter light out of any incident beam and in aggregate
are approximated by a smooth screen with an optical transmission of
$\taur$ that varies as a function of radial distance $r$ from the centre
of the secondary companion.
The rings are assumed to be azimuthally symmetric and the inclination
of the ring plane as seen from the Earth is $\id$ (with $0^o$ being
face-on) and the projected angle between the normal of the secondary
companion's orbit and the normal of the ring plane is $\pd$.
The obliquity $\varepsilon$ of the ring plane is related to these two
angles by:

$\cos \varepsilon = \sin \id \cos \pd$

The rings are assumed to be considerably thinner than their diameter
i.e. a ``thin ring'' approximation.
In Saturn's ring system, thicknesses from tens of metres
down to an instrument resolved limit of tens of centimetres
\citep{Tiscareno13} have been observed.
Pole on, the rings form a concentric set of circles centered on the
secondary companion.
The light curve of a {\em point source} passing behind the ring
structure along an arbitrary chord is therefore symmetric in time
about the point of closest approach of the projected star position to
the companion.
What may not be so obvious is that a thin ring system tilted at an
arbitrary inclination will {\em also} produce a symmetric light curve,
regardless of the chord chosen.
This can be seen when one considers that a set of concentric ellipses
can be transformed into a set of concentric circles by a single shear
transformation whose shear axis is parallel to the chord.

\begin{figure*}
\centering
\includegraphics[width=\hsize]{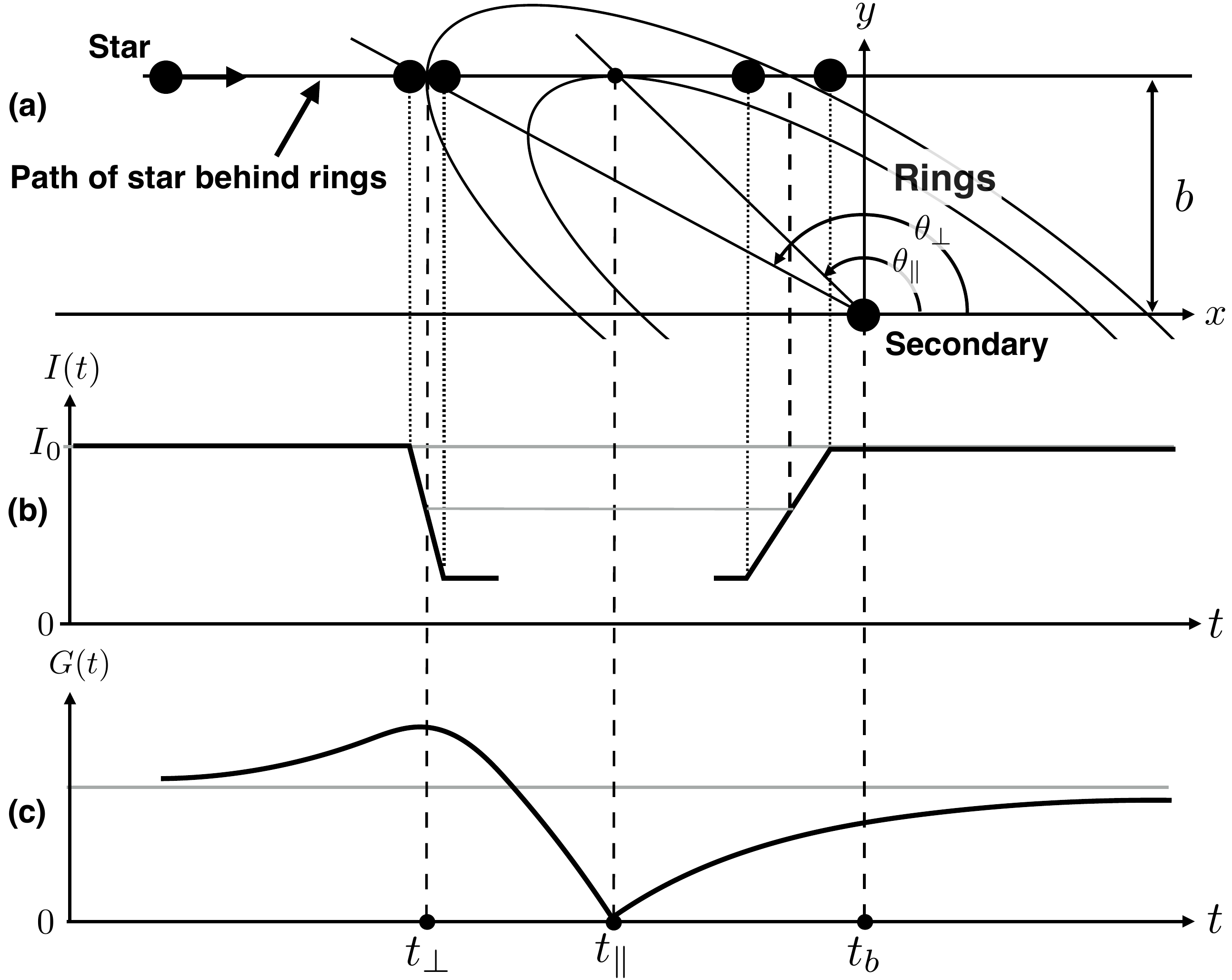}

\caption{The geometry of the ring model. Panel (a) shows a ring system
  inclined at an angle of $\id$ and rotated from the line of relative
  velocity by $\pd$.
The star passes behind the ring system with impact parameter $b$ at
time $t_b$. Panel (b) shows the resultant light curve $I(t)$ of the
star as a function of time, demonstrating how the local ring tangent
convolved with the finite sized disk of the star produces light curves
with different local slopes.
Panel (c) highlights the three significant epochs in the rate of
change of ring radius $r$; $t_b$ at closest projected separation of
the star and the secondary, $\tb$ where the ring tangent is
perpendicular to the direction of stellar motion, and $\tp$ where
stellar motion is tangent to the ring.
$\tp$ also marks where the stellar path touches the smallest ring
radius.\label{fig:ringgeom}}
\end{figure*}

The light curve $I(t)$ of a source with finite angular size (i.e. the
stellar disk of the primary) behind a tilted ring system, however, is
{\em not} time symmetric (see Figure \ref{fig:ringgeom}).
For each ring boundary, the gradient of the light curve $g(t)$ is
dependent on both the size of the star, and angle between the local
tangent of the ring edge and the direction of motion.
Ring structures smaller than that of the stellar diameter are smeared
out by the resultant convolution with the stellar disk, resulting in a
characteristic timescale defined by the time taken for the stellar
disk to cross its own diameter $t_\star=2R_\star/v$.

The track of the star on the projected ring plane has its closest
approach at time $t_b$ with an impact parameter of $b$, along with the
ring orientation defined by $\id$ and $\pd$.
$\id$ is the inclination of the plane of the rings to the plane of the
sky.
$\pd$ is the rotation of the ring system in the plane of the sky in an
anticlockwise direction.
These four parameters uniquely define the relationship between epoch
of observation $t$ and ring radius $r$.
We model an azimuthally symmetric ring of radius $r$ seen in
projection by an ellipse with the secondary companion at the origin
with a semi-major axis $r$ and semi-minor axis $r \cos i$.
The semi-major axis of the ellipse is rotated anticlockwise from the
x-axis by an angle $\phi$.
The parametric equation for a projected ring is then:

$$x(p) = r(\cos p \cos \pd - \sin p \cos \id \sin \pd)$$

$$y(p) = r(\cos p \sin \pd + \sin p \cos \id \cos \pd)$$

where $x$ and $y$ are coordinates on the ring at radius $r$ at a given
value of the parametric variable $p$ that has a value from $0$ to
$2\pi$ radians.

The star moves along a line parallel to the x-axis:

$$
\begin{array}{rl}
 x_\star & = v(t - t_b) \\
 y_\star & = b\\
\end{array}
$$

In addition to the time of closest projected approach of the star to
the secondary companion, there are two other significant epochs - the
epoch when the stellar motion is perpendicular to the ring-projected
ellipse $\tb$, and the epoch $\tp$ when the stellar motion is
tangential to the ring-projected ellipse.

To find where the tangent of the ring is perpendicular to the y-axis,
we see where $dx/dp=0$.  The result is then:

$$ \tan p_\bot = -\cos \id \tan \pd $$

Since nested rings only differ in a single scale factor centered on
the origin, the loci of all perpendicular tangents lie on a straight
line passing through the origin, i.e. the geometric centre of the
rings.
The angle $\theta_\bot$ can be calculated by substitution:

$$ \tan \theta_\bot = \frac{y(p_\bot)}{x(p_\bot)} $$

A similar derivation gives the angle for the line passing through the
loci of all parallel tangents, $\theta_\parallel$:

$$ \tan p_\parallel = (\cos \id \tan \pd)^{-1} $$

and

$$ \tan \theta_\parallel = \frac{y(p_\parallel)}{x(p_\parallel)} $$

Finally, for distances where the star/secondary companion distance is
much larger than the impact parameter (i.e where $x$ tends to very
large values with $y=0$) the gradient $dy/dx$ tends towards an
asymptote defined by:

$$ \tan \theta_{y=0} = \left. \frac{dy}{dx} \right |_{y=0} = \frac{2(1 +
\cot^2 \pd \cos^2 \id)}{\sin 2\pd} $$

These three regimes are shown in the lower panel of Figure
\ref{fig:ringgeom} as the time of largest gradient, the gradient
touching at zero, and the asymptotic gradient at large positive and
negative values along the x-axis.

A simple ring model is uniquely defined with four numbers - the
orientation of the ring system (the inclination and obliquity), the
impact parameter $b$ and the epoch of closest projected approach to
the secondary companion $t_b$.
The transmitted intensity of light through a ring at radius $r$ with
$\taur$ is:

$I(r) = I_0 e^{-\taur}$

Since we do not know if the rings are a single thin screen of
particles or are optically thick, we do not correct $\tau$ for the
inclination of the ring system.

The greatest uncertainty in the model fitting is the diameter of the
star.
We therefore express the size of the rings in units of time,
converting back to linear sizes at the end of the modeling.
With an assumed stellar size and relative velocity, the system can be
converted back into units of length by multiplying by $v$.
A light curve model for a given star is produced with the diameter of
the star, its relative velocity with respect to the secondary
companion, the limb darkening parameter, the orientation and the
radial transmission of the ring system.

\section{The light curve of J1407}
\label{sec:j1407}

The complex light curve of J1407 was first discussed in
\citet{Mamajek12}, where intensity fluctuations of up to 95\% were
seen over a 56 day period in April and May 2007 towards this young
($\sim 16$ Myr) K5 pre-main sequence star in photometry taken as part
of the SuperWASP Survey \citep{Pollacco06,Butters10}.
After ruling out other simpler astrophysical explanations,
\citet{Mamajek12} concluded that the light curve was due to the
transit of a giant ring system orbiting an unseen secondary companion,
and that this ring system was considerably larger than the diameter of
the central star.
The star was simultaneously observed by three of the eight cameras in
the SuperWASP South array, and due to its location in the corner of
the field of view of these three cameras, there is a clear systematic
offset of 0.3 magnitudes seen between the cameras in the standard data
reduction photometry.
A dedicated reprocessing of all the raw photometric data successfully
removes both the systematic offsets seen in the light curves and also
the much smaller amplitude stellar variability \citep{vanWerkhoven14}.
We use the cleaned photometric data \citep{vanWerkhoven14} as the input
for our ring fitting model.
The observations over the 54 day period are not continuous but are
interrupted by the diurnal cycle and cloud cover at the observing site,
resulting in a completeness of 11.3\%.

Fitting is performed in a two step process - we first constrain the
orientation of the ring system using the gradients measured from the
light curve (Section \ref{sec:gradfit}), and then use these parameters
to generate a model of the ring transmission as a function of radius
from the secondary companion (Section \ref{sec:ringfit}).

The angular diameter of the star is $65.2\pm9.3\,\mu$arcsec, based on
a distance $d=133\pm 12$ pc and radius of $0.99\pm 0.11 R_\odot$
\citep{vanWerkhoven14,Kenworthy15}.
Treating a transiting ring as a semi-infinite knife edge, and assuming a
point source behind the rings, the angular separation between the
geometric edge of the ring shadow and the first diffraction maximum is
$1.22 \surd{\lambda/2d}$, giving an angular fringe separation at
$62\pm4\,$ nanoarcsec, about 1000 times smaller than the angular
diameter of the star.
Using geometric shadows is therefore a valid approximation for the model.

\subsection{Fitting the ring orientation using light curve gradients}
\label{sec:gradfit}

For a given set of ring orientation parameters $\id,\pd,b,t_b$, we can
determine the radial distances of the rings from the secondary companion
(i.e. the ring radius) at any epoch, $r(t)=f(\id,\pd,b,t_b,t)$, and also
determine $dr(t)/dt$.
The transmission of the disk as a function of $r$ is given by $\taur$.
Together with a model of the stellar disk that includes limb darkening
\citep[see ][]{vanWerkhoven14} and the functional form of $\taur$, we
can calculate the light curve of a ring system model for any epoch.
In practice, we calculate a grid of values of $r$ for the track of the
star behind the ring system, with a spatial resolution set by the
diameter of the star.
The transmission $I(r)$ is calculated for all points in the grid along
the track.
This grid of flux values is then convolved with a model of the stellar
disk, and the line of pixels along the impact parameter $b$ then
represents the measured flux $I(t)$.
We use 25 pixels across the diameter of the star to sample the limb
darkening and stellar disk.

\begin{figure*}
\centering
\includegraphics[width=\hsize]{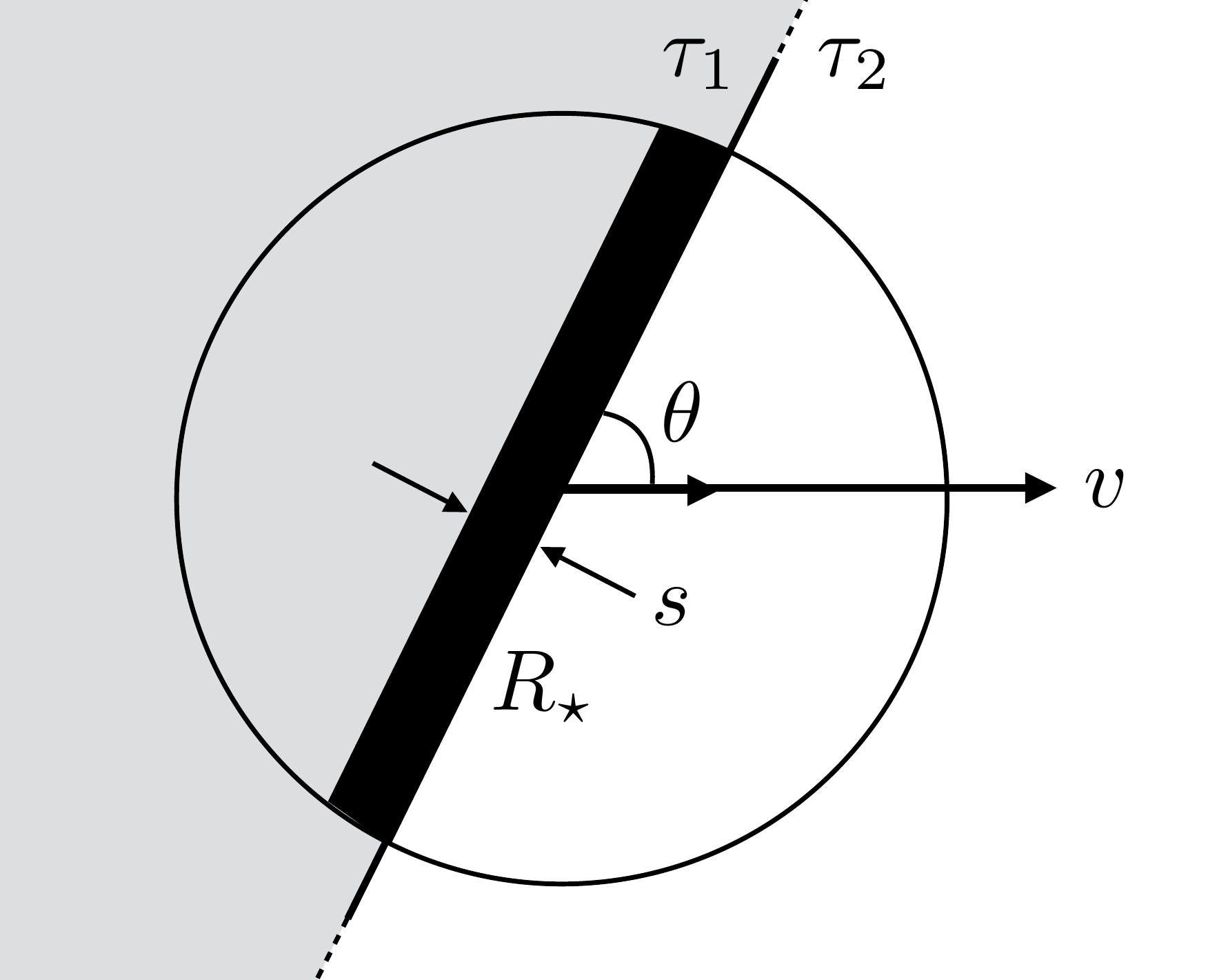}
\caption{Geometry of a ring edge crossing a stellar disk of uniform
intensity. The stellar disk is of uniform illumination with radius
$R_\star$ and we do not show limb darkening in this example.
The ring edge moves at a velocity $v$ across the disk. The ring edge is
at an angle $\theta$ to the direction of motion $v$. The black strip
has a width of $s$. The two rings have absorption coefficients of
$\tau_1$ and $\tau_2$.
\label{fig:edgedisk}}
\end{figure*}

\begin{figure*}
\centering
\includegraphics[width=\hsize]{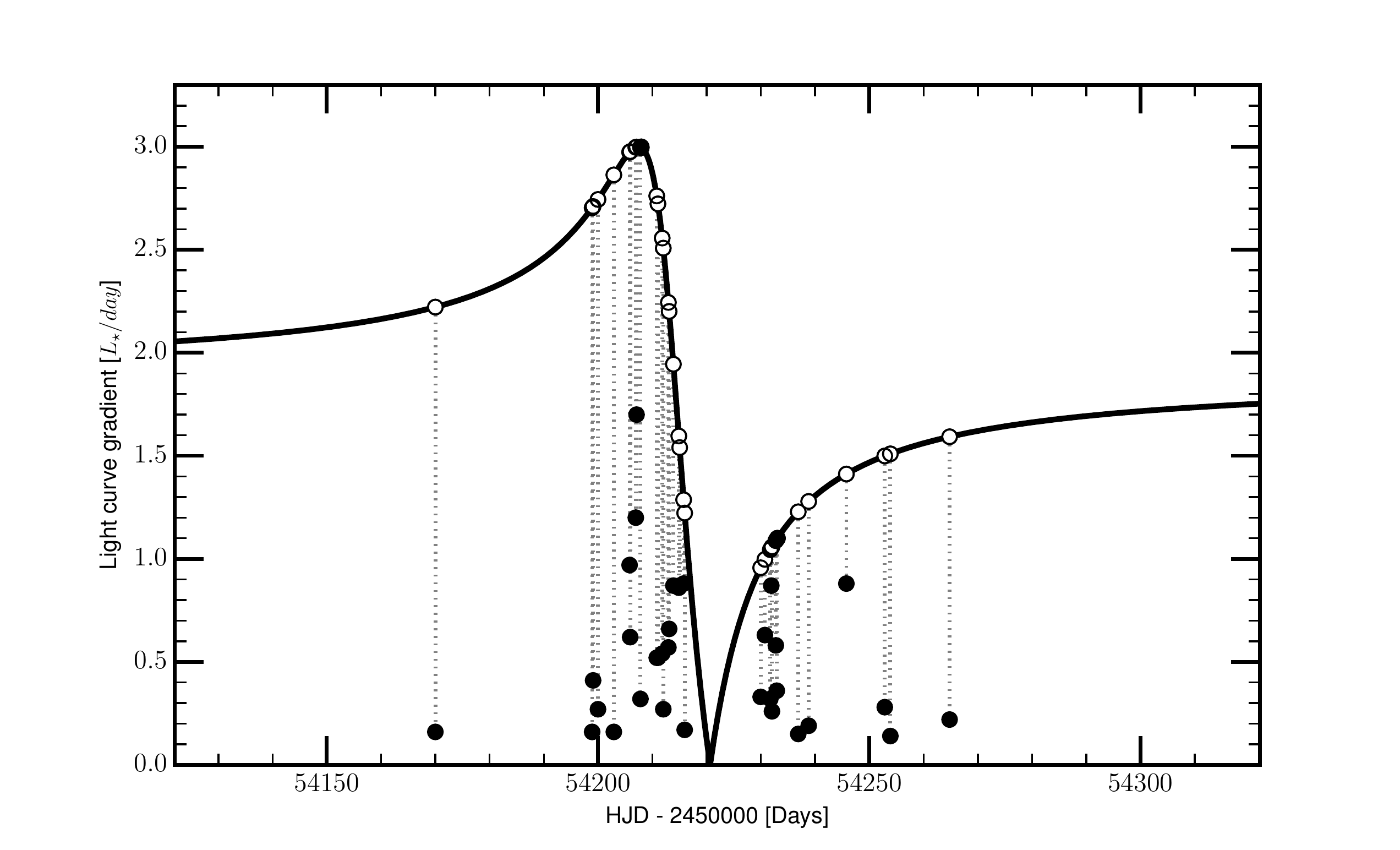}

\caption{The measured gradients in the light curve of J1407 plotted as a function of MJD of observation.
The line shows the maximum allowed gradient of the light curve $G(t)$ for a
given set of disk parameters $\tp$, $\tb$, $\id$ and $\pd$.
Dotted lines connect the black circle measured values to the maximum
allowed open circle theoretical maximum values on the black line.
\label{fig:diskfit}}

\end{figure*}

The measured flux $I(t)$ of J1407 can be closely approximated as a
sequence of straight lines of different gradients \citep[see ][ for
details]{vanWerkhoven14}.
We interpret these straight line light curves as a ring edge (between
two rings with different values of transmission) passing across the disk
of the star.
When the slope of the light curve changes, this represents a ring edge
either starting or finishing its transit of the stellar disk.
The measured gradients of straight line fits to the J1407 light curve
are shown as the black circles in Figure \ref{fig:diskfit}.
We assume that all the rings have well defined edges and have a constant
transmission across the width of the ring (with some assumed constant
$\tau$) - see Figure \ref{fig:edgedisk}.

To understand our ring orientation algorithm, consider a ring system
made up of alternately transparent and opaque rings whose radial width
would allow complete obscuration or transmission of the stellar disk.
The transmitted intensity goes from $I= I_0$ to $I=0$ and vice versa at
a rate determined by the ring velocity $v$ and the local tangent of the
ring to the line of stellar motion, defined as parallel to the x-axis in
our model.
If the gradient of the light curve is measured close to the midpoint of
the transit of a ring edge, and this quantity is plotted as a function of time,
the result is the black curve in Figure \ref{fig:diskfit}.

Defining the angle between the tangent of the ring at point $(x_\star,y_\star)$ to
the x-axis as $\theta$ (see Figure \ref{fig:edgedisk}):

$$ \tan \theta = dy/dx $$

The area swept out by a straight edge in time $dt$ across a stellar disk
with no limb darkening is equal to $2R_\star
s$ where $s= v dt \sin \theta$.
Defining the transmission $T_n = I_n/I_0
= \exp(-\tau_n)$, the change in intensity $dI$ is simply the change in
transmission from $T_1$ to $T_2$ over the area $2R_\star v \sin \theta
dt$, so the rate of change of intensity is:

\begin{equation}
g(t) = \frac{dI(t)}{dt} = (T_1 - T_2) G(t) = (T_1 - T_2) \frac{2v \sin
\theta}{\pi R_\star} \left ( \frac{12-12u+3\pi u}{12-4u} \right )
\label{gt}
\end{equation}

The limb darkening of the star is parameterised by $u=0.8(5)$ for J1407
\citep{Claret11}.

The function $G(t)$ represents the maximum flux change possible between a
fully transparent and fully opaque ring.
For rings that have intermediate values of transmission, the resultant
light gradients will lie underneath this black curve, and so the curve
represents an upper bound on the light gradient for a given ring
orientation.
Since we do not know $\taur$, we can use $G(t)$ as an upper limit
and we search for ring orientations that have all measured gradients lie
underneath this curve.

We have $n$ measurements of the light curve gradient $g(t)$ at time $t$.
The model gradient $G(t)$ is calculated from Equation \ref{gt}, where $\theta$ is a function of $\id,\pd,b,t_b,t$.
We calculate a cost function that minimises the difference between the
model and measured gradients, and penalizes heavily if the measured
point goes above the model point.
If we define $\delta_t = G(t) - g(t)$, then our cost function $\Delta$
is:

$$
\Delta = \sum_{t=1}^n \left\{ \begin{array}{rl}
\delta_t &\mbox{ if $\delta_t>0$} \\
-50 * \delta_t &\mbox{ otherwise}
\end{array} \right.
$$

The factor of 50 used in the above equation reflects the error of 2\% on
the measured light curve gradients, and prevents fitting algorithms from
oscillating near local minima.
We use an Amoeba simplex algorithm \citep{Press92} to solve for the four free
parameters.
The resulting behaviour of the cost function is to bring down the model
curve $G(t)$ so that at least two measured gradient points from $g(t)$
lie on $G(t)$, which may not be necessarily correct if the data is
sparse and the measured slopes do not sample the largest gradient of
$G(t)$ at $\tb$.
In the case of J1407, the largest observed gradient is during MJD 54220,
and the complete set of parameters for the disk geometry is listed in
Table \ref{tab:model1}.
It is highly probable that this is the largest gradient in the ring
system since the secondary companion velocity derived from this gradient
presents a challenge to the orbital dynamics for rings \citep[see ][ for
a detained discussion]{vanWerkhoven14,Kenworthy15}.
We therefore introduce an additional cost function that fixes the
midpoint of the eclipse light curve $\tb$ to a user defined value, so that
we can explore different ring geometries that still produce reasonable
cost functions.

\capstartfalse

\begin{deluxetable}{lcccccc}
\tablewidth{0pt}
\tablecaption{Disk model parameters\label{tab:model1}}
\tablehead{\colhead{Model} & \colhead{$b$} & \colhead{$t_b$} & \colhead{$i_{disk}$} & \colhead{$\phi_{disk}$} & \colhead{$t_\parallel$} & \colhead{$v$} \\
\colhead{} & \colhead{(d)} & \colhead{(d)} & \colhead{(deg)} & \colhead{(deg)} & \colhead{(d)} & \colhead{km.s$^{-1}$} }
\startdata
1 & 3.92 & 54225.46 & 70.0 & 166.1 & 54220.65 & 33.0 \\
\enddata
\end{deluxetable}

\capstarttrue

The minimum velocity required to cross a limb-darkened star is derived
in \citet{vanWerkhoven14}, and for the case of J1407, Equation 12 gives
a relative velocity of $33\mathrm{km.s}^{-1}$ for $R_\star=0.99 R_\odot$
and $\dot{L}_{max}=3.1 \mathrm{L}_*/\mathrm{day}$.
We adopt these values for our model.

\subsection{Fitting the ring structure}
\label{sec:ringfit}

The ring radius $r(t)$ can be calculated with values for $\id,\pd,b,t_b$
estimated from the disk fitting procedure of the previous section.
We now look for the ring transmission as a function of radius
by using our model of $r(t)$, the stellar radius and limb darkening
profile of J1407, and an estimate of the transverse velocity $v$.
We adopt the limb darkening profile and parameters of
\citet{vanWerkhoven14} and a stellar radius of $0.99{\rm R}_\odot$
\citep{Kenworthy15}.
The number of ring edges in the light curve are estimated by counting
the number of slope changes identified in the light curve and
indirectly implied by the change of the light curve during daylight
hours.
At least 24 ring edges are required for the number of gradient changes
detected in the J1407 data \citep{vanWerkhoven14}, but given the
sparseness of the photometric coverage this number is almost certainly
higher.

Using the derived ring geometry parameters, the absolute value of the
time since the closest approach of the secondary companion,
abs$(t-\tp)$, is used as the origin for a graph that displays the
observed light curve, the model light curve, and the difference of these
two curves.
Displaying the data and model in this way allows a direct visual
comparison between the ingress and the egress of the star about the time
of closest approach of the secondary companion $\tp$.
In the limiting case of a point-like background source, the ingress
light curve and egress light curve are identical for azimuthally
symmetric ring systems.
The finite diameter of the star breaks this degeneracy and so the
ingress and egress light curve for a given ring edge radius are not
identical.
We generate an initial estimate of $\taur$ using a GUI written in the
Python programming language.
This is defined as:

\begin{equation} \label{eq:ring} \taur = \tau_n \mathrm{\ for\ } r_{n-1}
< r < r_n \mathrm{\ with\ } r_0=0 \mathrm{\ and\ }n =
1\mathrm{\ to\ }N_{rings} \end{equation}

Ring edges and transmission values are interactively added, moved and
deleted as appropriate and the ring model $I(t)$ is generated after each
successive operation.
A visual inspection of the model light curve and its comparison to the
data is carried out, and when the minimum number of rings are added to
the model, the ring transmission values are optimized using a minimum
least squares fit to the data and Amoeba simplex algorithm.

There are 16489 photometric data points covering the 2007 observing
season of J1407 (MJD 54131.96 to 54306.72).
The data of J1407 does not show significant changes in flux photometry
over the timescale of 30 minutes, and so to speed up the interactive
fitting, the photometric data is re-binned in 0.02 day (32 minute)
intervals with error estimates calculated from the r.m.s. of the
photometric data points within each bin.
Bins containing less than three photometric points are discarded,
resulting in a data series of 985 points.
The photometric data is sparse (the binned data covers 11.3\% of the
total 2007 season), and so no unique solutions for the disk geometry and
$\taur$ are found.\footnote{When the light curve is folded at time of
closest approach, the coverage approximately doubles to 20\%.
This implies that there are at least $24/0.2\approx 120$ ring edges
in the J1407b system.}

\capstartfalse

\begin{deluxetable}{ll}
\tablewidth{0pt} \tabletypesize{\scriptsize}
\tablecaption{Table of ring parameters \label{tab:ring}}
\tablehead{\colhead{Ring Outer Edge Radius} & \colhead{Tau} \\
\colhead{($10^6$ km)} & \colhead{}}
\startdata
30.8 & 4.65 \\
31.3 & 0.93 \\
32.5 & 2.18 \\
33.0 & 0.52 \\
34.0 & 1.47 \\
35.4 & 3.61 \\
36.2 & 1.01 \\
37.4 & 0.24 \\
38.0 & 1.07 \\
39.1 & 0.21 \\
40.6 & 0.69 \\
42.3 & 0.40 \\
42.6 & 1.01 \\
43.5 & 0.37 \\
46.6 & 0.72 \\
48.0 & 1.53 \\
49.5 & 0.38 \\
50.4 & 2.66 \\
51.2 & 0.03 \\
51.7 & 1.54 \\
52.9 & 0.80 \\
53.5 & 0.00 \\
53.9 & 2.94 \\
55.3 & 0.70 \\
57.2 & 0.30 \\
59.2 & 0.59 \\
61.2 & 0.00 \\
63.0 & 0.50 \\
65.5 & 0.21 \\
66.7 & 0.11 \\
68.9 & 0.12 \\
75.4 & 0.08 \\
78.5 & 0.25 \\
83.0 & 0.06 \\
90.2 & 0.52 \\
\enddata
\end{deluxetable}

\capstarttrue

\begin{figure*}
\centering
\includegraphics[angle=0,width=16cm]{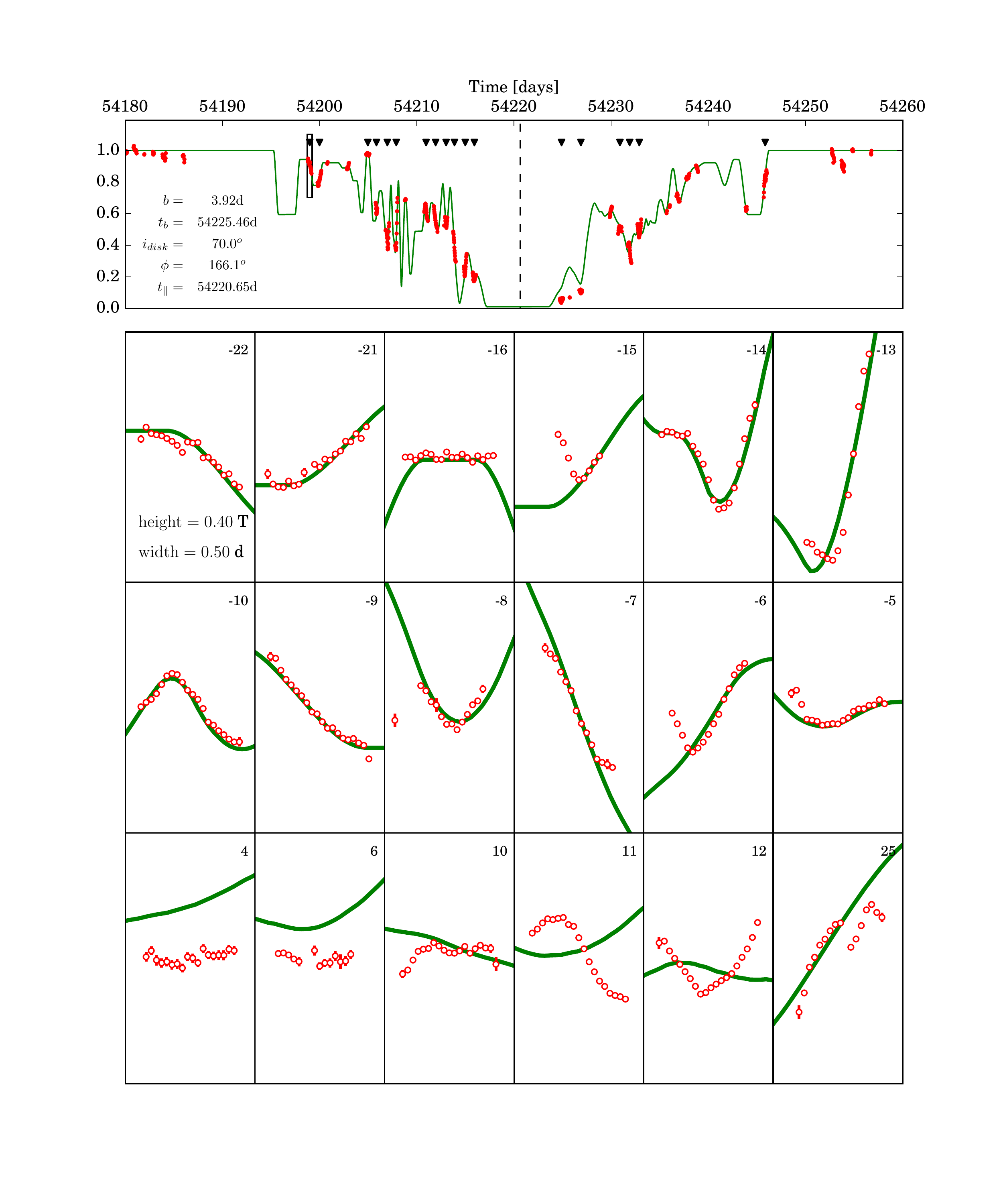}

\caption{Photometry of J1407.
The upper panel shows the light curve of J1407 as the red points with
associate error bars.
The green curve is the model fit for Model 1, with
$t_\parallel=54220.65$ d (indicated with the vertical dashed line) and
$v=33$ km.s$^{-1}$.
For the nights indicated with the inverted triangles, the photometry and
model fit is enlarged into the panels below.
Each panel has a width of 0.5 days and a height of 0.4 in transmission.
The number in the top right hand corner represents the number of days
from $t_\parallel$.  \label{fig:dailyfit}}

\end{figure*}

\begin{figure*}
\centering
\includegraphics[angle=90,width=12cm]{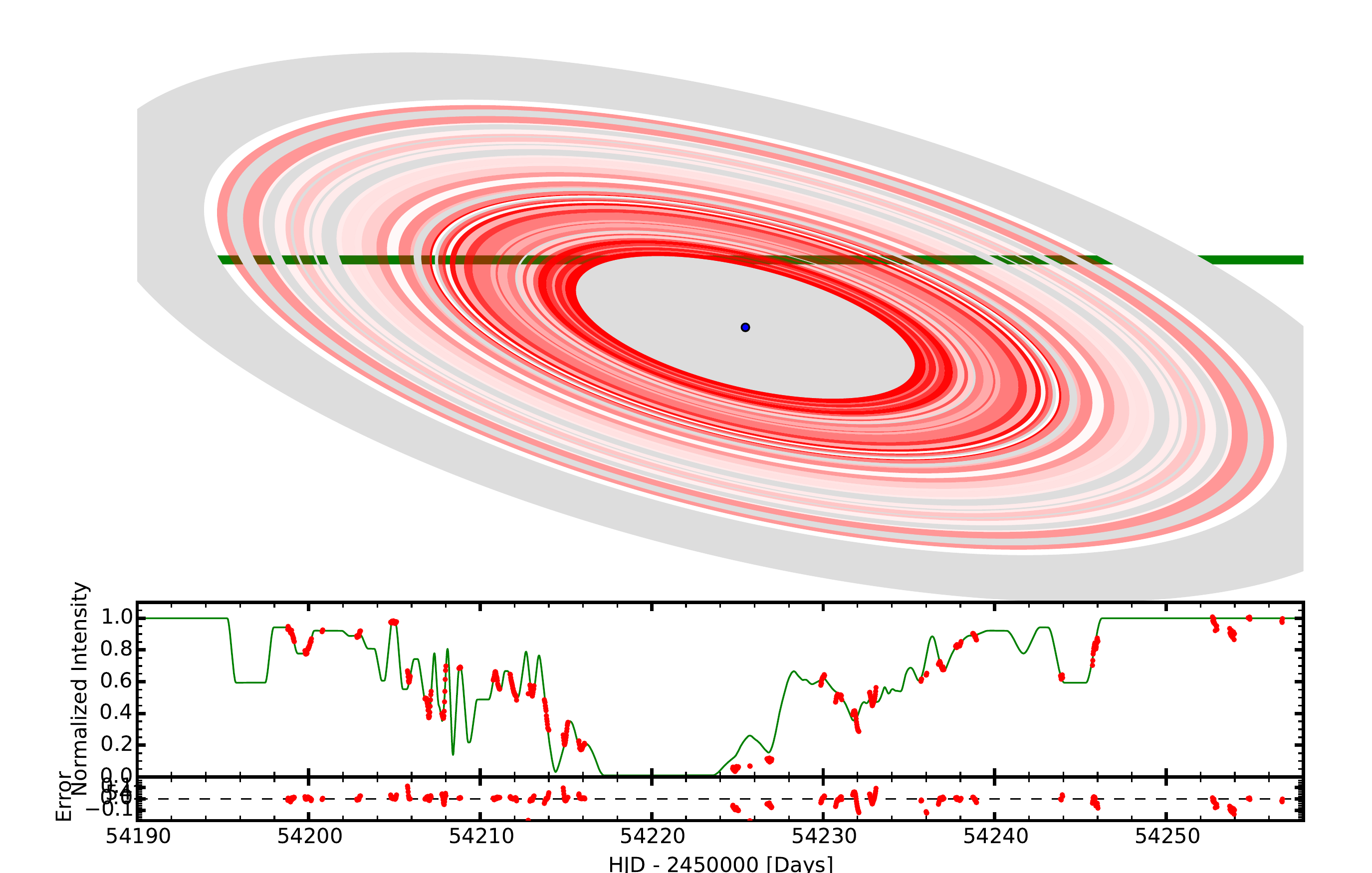}

\caption{Model ring fit to J1407 data.
The image of the ring system around J1407b is shown as a series of
nested red rings.
The intensity of the colour corresponds to the transmission of the ring.
The green line shows the path and diameter of the star J1407 behind the
ring system.
The grey rings denote where no photometric data constrain the model fit.
The lower graph shows the model transmitted intensity $I(t)$ as a
function of HJD.
The red points are the binned measured flux from J1407 normalised to
unity outside the eclipse.
Error bars in the photometry are shown as vertical red bars.
\label{fig:taufit}}

\end{figure*}

In Figures \ref{fig:dailyfit} and \ref{fig:taufit} we present one
possible ring solution (listed as Model 1 in Table \ref{tab:model1}) to
the J1407 photometric data, where the central eclipse is set at $t_b =
54220.65$ MJD.

\section{Interpreting the J1407b Model}
\label{sec:structure}

By visual examination there is a clear decrease and subsequent increase
in the transmission of J1407 flux with a minimum around MJD 54220.
Using photometry averaged in 24 hour bins, a ring model with four broad
rings is consistent with data on these timescales \citep{Mamajek12}.
On hourly timescales, the large flux variations are consistent with
sharp edged rings crossing over the unresolved stellar disk.
We do not find an azimuthally symmetric ring model fit that is
consistent with all the photometric data at these timescales.
Due to the incomplete photometric coverage, there are several models
that fit with similar $\chi^2$ values to the data.
In all of these cases, we see the presence of rapid fluctuations in the
ring transmission both as a function of time and as a function of radial
separation from the secondary companion.

\begin{figure}
\centering
\includegraphics[width=\hsize]{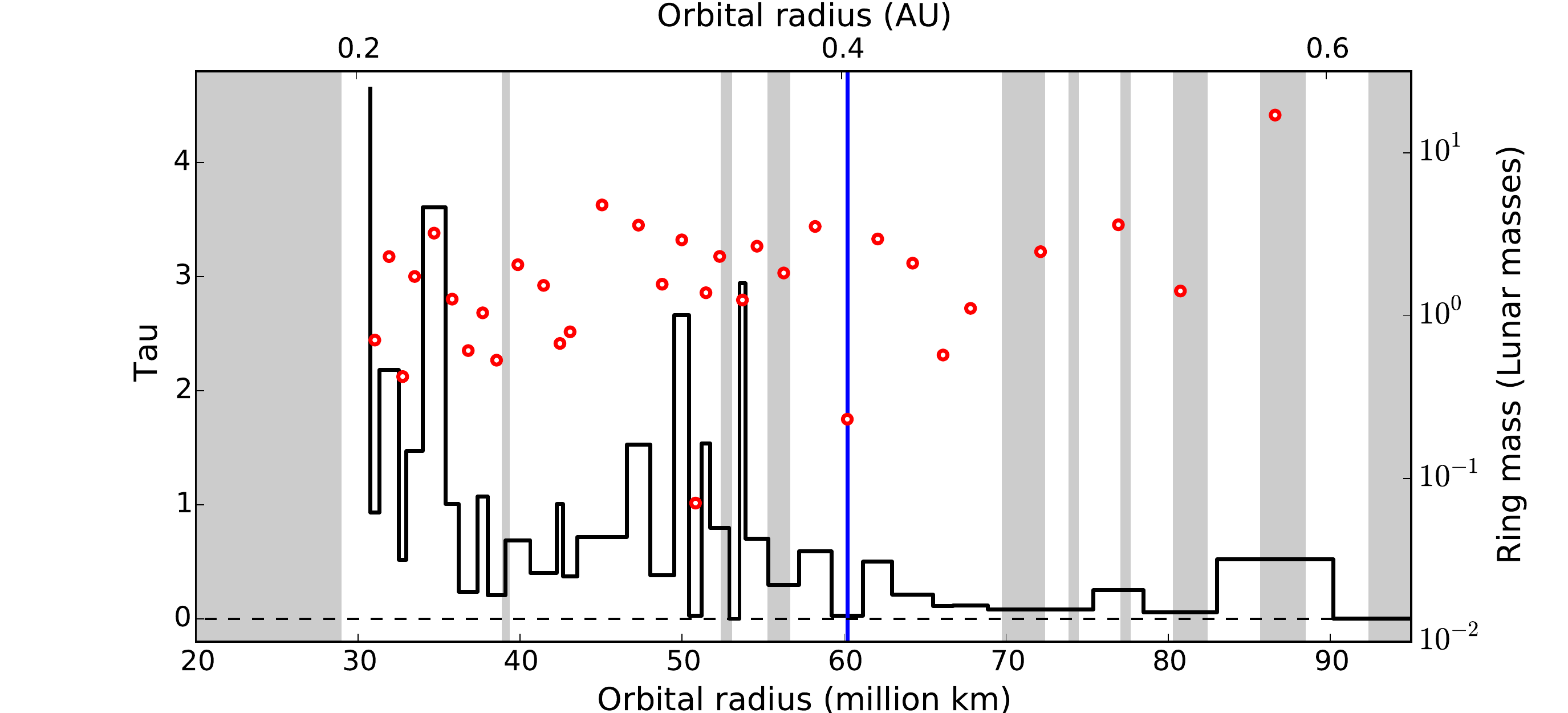}

\caption{The transmission of the ring model as a function of radius.
The grey regions indicate where there in no photometry to constrain the
model.
The blue line indicates the ring gap seen at 61 million km.
Red dots indicate the estimated mass of each ring assuming a mass
surface density of $\sim 50\, {\rm g}\ {\rm cm}^{-2}$.
\label{fig:tfit}}

\end{figure}

There are clear gaps in all the ring model solutions explored.
Gaps in the rings of Solar system giant planets are either caused
directly by the gravitational clearing of a satellite or indirectly by a
Lindblad resonance due to a satellite on a larger orbit.
The J1407 ring system is larger than its Roche limit for the secondary
companion.
A search for the secondary companion is detailed in \citet{Kenworthy15},
and the constraints from null detections in a variety of methods result
in a most probable mass and orbital period for the secondary companion.
These orbital parameters are summarised in Table \ref{tab:1407b_orbit}.
We take the most probable mass and period for the moderate range of
eccentricities with mass 23.8$M_{Jup}$ and orbital period $13.3$ yr,
although we note that the period could be as short as 10 years and the
mass can be greater than 80 $M_{Jup}$, but mass this is considered highly
unlikely with a probability of less than 1.2\%.
Gaps in the ring system are either seen directly as the photometric flux
from J1407 returning to full transmission during the eclipse, or
indirectly as a fit of the model to intermediate transmission
photometric gradients.
One ring gap with photometry is at HJD 54210, seen during the ingress of
J1407 behind the ring system.
The corresponding radius for this gap in the disk is seen from 59 million km to 63 million km
(indicated in Figure \ref{fig:tfit}), corresponding to an orbital period
$P_{sat}$ of:

$$ P_{sat} = 1.7\,\mathrm{yr}\left( \frac{M_{J1407b}}{23.8 M_{Jup}}
\right )^{-1/2} $$

If we assume that the gap is equal to the diameter of the Hill sphere of
a satellite orbiting around the secondary companion and clearing out the
ring, then an upper mass for a satellite can be calculated from:

$$ m_{sat} \approx 3 M_b \left ( \frac{d_{hill}}{2a} \right )^3 =
0.8 M_\oplus \left ( \frac{M_{J1407b}}{23.8 M_{Jup}} \right ) $$

For the case of $23.8M_{Jup}$ this corresponds to a satellite mass of
$0.8 M_\oplus$ and an orbital period of 1.7 yr.

\capstartfalse

\begin{deluxetable}{lll}
\tablewidth{0pt} \tabletypesize{\scriptsize}
\tablecaption{Table of J1407b orbital parameters \citep[from][]{Kenworthy15}  \label{tab:1407b_orbit}}
\tablehead{\colhead{Orbital Eccentricity} & \colhead{$0.7 < e < 0.9$} & \colhead{$0.7 < e < 0.8$} }
\startdata
Probable period $\overline{P}$ (yr)          & 27.5   &  13.3 \\
Probable mass $\overline{M}$ ($M_{Jup}$)     & 14.0   &  23.8 \\
\enddata
\end{deluxetable}

\capstarttrue

The ring transmission at smaller radii shows structures that are analagous
to the Kirkwood gaps in the Solar System, where the smooth radial
distribution of asteroids is interrupted by peturbations due
to period resonances with Jupiter.
In the case of the largest ring the Solar system around Saturn, the
Phoebe ring is not coplanar with the other rings \citep{Verbiscer09} but
lies in the plane of Saturn's orbit.
This is due to the dominance of solar perturbations over the
gravitational perturbation caused by the $J_2$ contribution of Saturn's
gravitational field.
The planet $\beta$ Pictoris b was recently shown to have a rotation
period of $\sim 8$ hours \citep{Snellen14}, faster than that of the
other gas giants in the Solar system.
The $\beta$ Pictoris system also has an age of $22$ Myr
\citep{Mamajek14}, similar to the J1407 system and implying that a low mass companion
could have a similar fast rotation period, larger oblateness resulting
in a larger $J_2$ contribution, holding ring structures in the
equatorial plane out to larger radii.

\subsection{Notable timescales in the giant exoring model}

There are two additional timescales that are worth noting in this giant
exoring model.
At the time of $\tp$, the star is close to stationary in $r(t)$ and is
moving tangentially to the ring at radius $r(\tp)$.
There is therefore a time interval $\Delta t_1$ where the star is
sensitive to variations in the azimuthal structure around radius
$r(\tp)$.
The length of time for this interaction is approximately the length of
time it takes for the projected disk radius to change by the diameter of
the star, i.e. where:

$$r(\tp) + (R_*/\cos \id) = r(\tp+\Delta t_1)$$

This timescale for the J1407 exoring system is approximately 1 day
around $\tp$.
We cannot examine the J1407b data for $\Delta t_1$ as there is no
recorded photometry within 3 days of $\tp$.

The second timescale relates to ring material orbiting around the
secondary companion and the duration of the disk transit.
If the secondary companion is large and the impact parameter
$b$ small enough, there will be a parcel of ring material that will
transit in front of the star at least twice.
The orbital motion of ring material about the secondary companion,
combined with the orbital motion of the secondary companion about the
primary star will result in two epochs symmetric about $\tp$, where the
measured transmission will be of the same parcel of ring material.
Tests for ring illumination geometry and dust scattering can then be
carried out, to name one possibility for this.
The timescale for J1407 is less than 4 days, with a more specific number
dependent on disk geometry and secondary mass of the companion.

\section{The Orbit of J1407b}
\label{sec:orbit}

Only one eclipse of J1407 is seen in the publically available
photometric data, and so we do
not know the orbital period of J1407b or even if it is bound to J1407
in a closed orbit\footnote{By using the name J1407b for the eclipsing object, we
are implicitly assuming that the ring system is bound to J1407 for the
reasons expanded on later in this section.}.
We explore two hypotheses for the motion of the ring system relative to the
star J1407: (i) the ring system is unbound to J1407 and is a
free-floating planet with a ring system and (ii) J1407b is in a bound
and closed orbit about J1407.
Investigations into the latter case are detailed in \citet{vanWerkhoven14}
and \citet{Kenworthy15} respectively.

One estimate of the transverse ring system velocity is calculated from
the diameter of the star J1407 and the steepest light curve gradient
seen in the photometric light curve data.
Assuming a sharp-edged opaque ring crossing the disk of the star with
the ring edge perpendicular to the direction of motion, a minimum
velocity of 33 km.s$^{-1}$ is derived \citep{Kenworthy15}.
Combined with the duration of the eclipses, an estimate of the size of
the ring system can be made.
It is this derived transverse velocity and associated ring system size
that forms our central issue with the nature of the system.

\subsection{The ring system is an unbound object}

We consider if the ring system is on an unbound trajectory with a tangential velocity of
at least 33 km.s$^{-1}$.
The May 2007 eclipse is therefore a single event that will not be repeated
again with J1407.
Our derived ring model is consistent with the data, yielding a diameter
of 1.2 AU for the ring system.
We consider the unbound hypothesis as exceptionally unlikely, for
two reasons:

(1) The mean projected separation between stars in the field is $\approx
10^3$ AU. The probability that a 1 AU scale object produces an eclipse within
the lifetime of SuperWASP is exceptionally small.

(2) J1407b is substellar and almost certainly planetary mass
\citep{Kenworthy15}.
The estimated size of the ring system is two orders of magnitude larger
than the Roche limit for the central substellar mass, and so the rings
outside the Roche limit are expected to accrete into satellites on
timescales shorter then Gigayears.
This implies that the ring system is considerably younger than stars
seen in the field.

Direct imaging limits reported in \citet{Kenworthy15} constrain such a
free-floating object to be $8 M_{Jup}$, assuming an age of 16 Myr and
BT-SETTL models \citep{Allard12}.
We conclude that the ring system is bound to J1407 in a closed orbit.

\subsection{J1407b is on a bound orbit}

In \citet{Kenworthy15} a search for J1407b is carried out using
photometry, radial velocity measurements of J1407, direct and
interferometric techniques.
The companion is not detected, but upper limits from these observations
constrain the possible orbital period and mass of J1407b.
A circular orbit for J1407b would mean that any ring system would not be
subject to gravitational peturbations due to the orbit of J1407b, and so
appears preferable for ring stability.
Transverse velocities derived from the light curve gradients, however,
strongly rule out long orbital periods, but the lack of a second primary
eclipse in time-series photometry rules out short orbital periods.
A circular orbit is possible if the rings are themselves clumpy in
nature and the clump orbital motion vectorially adds to the J1407b
orbital motion \citet{vanWerkhoven14}, but this again requires a series
of coincidences to occur to generate a large gradient at the appropriate
epochs.

An elliptical orbit with the eclipse coincident with periastron passage
of J1407b can provide the transverse velocity required for the ring
model.
Constraints presented in \citet{Kenworthy15} suggest a minimum
eccentricity of 0.7 and a minimum period of 10 years.
The longest period is unconstrained in this model, as very highly
eccentric orbits are possible with the long axis pointing towards Earth,
but the probability of such a precise alignment becomes increasingly unlikely.
Truncating the eccentricity to be between 0.7 and 0.8 gives a probable
mass of 24 $M_{Jup}$ and probable period of 13.3 years.
One conclusion with these eccentric orbits is that the ring
system is within the Hill sphere for most of its orbit, but at
periastron the Hill sphere shrinks down below the size of the outermost
rings, providing a challenge for investigations into the stability of
this ring system.

\section{Conclusions}
\label{sec:conc}

We interpret the 56 day J1407 light curve event centered on UT 30 April
2007 as being due to a highly structured ring system surrounding an
unseen secondary companion, supporting the conclusions in
\citet{Mamajek12,vanWerkhoven14,Kenworthy15}.
Using the gradients in the light curve generated by the finite size of
the primary star, we solve for the geometry and impact parameter of the
secondary companion and ring system.
The size of the ring system is considerably larger than the Roche radius
for the secondary companion, and fills a significant fraction of the
Hill radius.
This implies that this structure is in a transitional state, with the
rings at large radii undergoing accretion to form exosatellites orbiting
the secondary companion.
The rapid variation in ring density as a function of radius implies
dynamical clearing processes that keep material out of the ring
plane.
Two such processes are (i) the formation of exomoons that are
gravitationally clearing out these gaps, such as those sculpting the
A-ring gaps in Saturn, and (ii) the presence of Lindbad resonances, caused by
the presence of unseen exosatellites at larger radii.
For a secondary companion mass of $28 M_{Jup}$, we interpret one of the
most well-defined ring gaps
at 61 million km with a width of 4 million km to be cleared by a
satellite with an upper mass of $< 0.8 M_\oplus$.
This satellite would have an orbital period of $\sim 2$ years in the
ring system about J1407b.

We estimate the mass of the individual rings by assuming a dust opacity
of $\kappa \sim 0.02\, {\rm cm}^2\ {\rm g}^{-1}$ for unity optical
depth, as assumed in \citet{Mamajek12} for their estimate.
The mass of each model ring is calculated without any correction for the
projected inclination of the rings, and is then plotted as the open
circles in Figure \ref{fig:tfit}.
The mass of the rings is dependent on the orbital parameters of J1407b,
the effective dust opacity in the rings, the detailed dust size
distribution and the unknown ring structure where there is no
photometric constraint.
In the regions of no photometry, the ring structure is extended with the
same optical properties at that at the closest known edge.
The order of magnitude estimate for the total mass of the rings is
$\sim 100M_{Moon}$, close to the mass of the Earth.
The ratio of the mass of the densest rings around Saturn (the B ring
system) to the largest satellite, Titan, is approximately 1/4000.
A large fraction of the mass is therefore accreted in the satellites.
It is interesting to note that the satellite in J1407b has a similar
mass (to an order of magnitude) as the rings, implying that further
accretion into satellites is ongoing.
For the cases where there are significant gaps in the photometric
coverage or there is a window function imposed by the diurnal cycle,
degeneracies appear in the fitted exoring models.
Photometric data that continuously samples the eclipse light curve can
completely solve the geometry of the exoring system.

Ring systems are thought to occur around other exoplanets, although none
have been confirmed.
Fomalhaut b is a co-moving companion to the nearby 400 Myr old star,
moving on an eccentric orbit \citep{Kalas13,Mamajek12b}.
A large ring system around the planet is thought to cause the
anomalously bright flux seen in optical images.
The orbit for $\beta$ Pictoris b is close to edge on \citep{Nielsen14},
indicating that it might transit its parent star.
Anomalous photometry of $\beta$ Pictoris in 1981
\citep{LecavelierdesEtangs95} has also been attributed to an extended
system of material surrounding the planet, and the anomalous photometry
extends over 30 days, implying that there is material extending out to
$0.1$ of the Hill radius of the planet.
This could plausibly be caused by a giant ring system transiting $\beta$
Pictoris, similar to the ring system about J1407b.

With simple assumptions on ring geometry and the ring plane orientation,
this ring model reproduces many but not all of the nightly photometric light curves.
These discrepancies imply either an error in the determined geometry of
the ring plane, and/or the rings are not coplanar.
Further modeling with additional degrees of freedom for the rings, such
as warping and precession may lead to better fits to the photometric
data.

J1407 is currently being monitored both photometrically and
spectroscopically for the start of the next transit.
A second transit will enable a wide range of exoring science to be
carried out, from transmission spectroscopy of the material, through to
Doppler tomography that can resolve ring structure and stellar spot
structure significantly smaller than that of the diameter of the star.
The orbital period of J1407b is on the order of a decade or possibly
longer.
Searches for other occultation events are now being carried out
\citep{Quillen14} and searches through archival photographic plates
\citep[e.g. DASCH;][]{Grindlay12}, may well yield several more
transiting ring system candidates.

\acknowledgments

We would like to acknowledge the anonymous referee for their careful
reading of our paper and their comments which have improved it.
We have used data from the WASP public archive in this research. The
WASP consortium comprises the University of Cambridge, Keele University,
University of Leicester, The Open University, The Queen's University
Belfast, St. Andrews University, and the Isaac Newton Group. Funding for
WASP comes from the consortium universities and from the UK's Science
and Technology Facilities Council.
EEM acknowledges NSF award AST-1313029.

{\it Facilities:} \facility{SuperWASP}

\bibliographystyle{apj}

\end{document}